\documentclass[aps,prd,%multicol,twocolumn, 
preprintnumbers,groupedaddress]{revtex4}
\usepackage{amsmath}

\begin{document}
\preprint{\vtop{
{\hbox{YITP-13-31}\vskip-0pt
%                 \hbox{hep-ph/07????} 
}
}
}

%\date{\today}

\title{
Decay Property of Axial-Vector Tetra-Quark Mesons 
}

\author{
Kunihiko Terasaki   %authors' name%
}
\affiliation{
Yukawa Institute for Theoretical Physics, Kyoto University,
Kyoto 606-8502, Japan 
}

\begin{abstract}{
Decay property of hidden-charm tetra-quark mesons is studied. 
It is seen that estimated width of iso-triplet odd $\mathcal{C}$ partners of $X(3872)$,
although still crude, is compatible with the measured ones of $Z^{\pm,0}_c(3900)$. 
It is pointed out that confirmation of $\hat{\delta}^{c0}(3200)$ (an $\eta\pi^0$ peak  
around 3.2 GeV indicated in $\gamma\gamma$ collision) gives a clue to select a 
realistic model of multi-quark mesons.  
}
\end{abstract}

\maketitle

Recently charged hidden-charm mesons $Z_c^\pm(3900)$ have been discovered in  
$\pi^\pm J/\psi$ channels of 
$e^+e^-\rightarrow Y(4260)\rightarrow \pi^+\pi^-J/\psi$~\cite{BESIII},  
and just after the observation, $Z^\pm(3895)$ have been observed in the same type of  
reaction~\cite{Belle-Z(3895)}, and then, not only $Z_c^\pm(3900)$ but also a neutral  
$Z_c^0(3900)$ has been found in data on 
$e^+e^-\rightarrow \psi(4160)\rightarrow \pi^+\pi^-J/\psi$  
and $\pi^0\pi^0J/\psi$ from CLEO-c~\cite{Xiao}. 
When $Z_c^\pm(3900)$ and $Z^\pm(3895)$ are identified, the average values of their  
masses and widths are given by 
%%%%%%%%%%%%%%%%%%%%%%%%%%%%%%%%%%%%%%%%%%%%%%%%%%%%%%%%%%%%%%%%%%%%%%%%
$\langle{m_{Z_c^\pm(3900)}}\rangle = 3894\pm 5$ MeV 
and $\langle{\Gamma_{Z_c^\pm(3900)}}\rangle = 48\pm 20$ MeV. 
%%%%%%%%%%%%%%%%%%%%%%%%%%%%%%%%%%%%%%%%%%%%%%%%%%%%%%%%%%%%%%%%%%%%%%%%
Regarding $Z^0_c(3900)$, however, its observation has been reported only in \cite{Xiao}, 
and its mass and width have been provided as 
%%%%%%%%%%%%%%%%%%%%%%%%%%%%%%%%%%%%%%%%%%%%%%%%%%%%%%%%%%%%%%%%%%%%%%%%
$m(Z^0_c(3900))=(3907\pm 12)$ MeV and 
$\Gamma(Z^0_c(3900))=(34\pm 29)$ MeV. 
%%%%%%%%%%%%%%%%%%%%%%%%%%%%%%%%%%%%%%%%%%%%%%%%%%%%%%%%%%%%%%%%%%%%%%%%
($J/\psi$ is written as $\psi$ hereafter.) 
In addition, existence of an $\eta\pi^0$ peak around 3.2 GeV (called as 
$\hat{\delta}^{c0}(3200)$ in this short note) 
%which is assigned to  $\hat{\delta}^{c0}$ in our earlier work~\cite{hidden-charm-scalar-KT}) 
has been indicated in $\gamma\gamma$ collision~\cite{Uehara}. 

On the other hand, tetra-quark models~\cite{Maiani,X-3872-KT} predicted existence of  
iso-triplet ($I=1$) partners of $X(3872)$ with opposite charge-conjugation ($\mathcal{C}$) property before the above observations of $Z^{\pm,0}_c(3900)$, and, 
after the observations, the prediction from the diquark-antidiquark model has been  
updated~\cite{Maiani-Z_c},  and these mesons have been newly interpreted as $I=1$  
opposite $\mathcal{C}$ partners of $X(3872)$ from different 
pictures~\cite{Z_c-mol,Voloshin}. 
Regarding $\hat{\delta}^{c0}(3200)$, it can be considered as the neutral component of  
hidden-charm $I=1$ scalar mesons~\cite{hidden-charm-scalar-KT}. 
These mesons as well as the well-established~\cite{PDG13} $D_{s0}^+(2317)$ and 
$X(3872)$ are considered as tetra-quark mesons from their decay properties, as seen  
below. 
The ratio of decay rates 
%%%%%%%%%%%%%%%%%%%%%%%%%%%%%%%%%%%%%%%%%%%%%%%%%%%%%%%%%%%%%%%%%%%%%%%
$R({D_s^{*+}\gamma/D_s^+\pi^0}) 
= {\Gamma(D_{s0}^+(2317)\rightarrow D_s^{*+}\gamma)}/
{\Gamma(D_{s0}^+(2317)\rightarrow D_s^+\pi^0)}$  
%%%%%%%%%%%%%%%%%%%%%%%%%%%%%%%%%%%%%%%%%%%%%%%%%%%%%%%%%%%%%%%%%%%%%%%
is experimentally constrained as~\cite{PDG13} 
%%%%%%%%%%%%%%%%%%%%%%%%%%%%%%%%%%%%%%%%%%%%%%%%%%%%%%%%%%%%%%%%%%%%%%%
$R({D_s^{*+}\gamma/D_s^+\pi^0})_{\rm exp}  < 0.059$.    
%%%%%%%%%%%%%%%%%%%%%%%%%%%%%%%%%%%%%%%%%%%%%%%%%%%%%%%%%%%%%%%%%%%%%%%
From this fact, it is natural to consider that $D_{s0}^+(2317)$ is a member of  $I=1$  
states, because of the well-known hierarchy of hadron interactions~\cite{HT-isospin}, 
%%%%%%%%%%%%%%%%%%%%%%%%%%%%%%%%%%%%%%%%%%%%%%%%%%%%%%%%%%%%%%%%%%%%%%%
$|${\it isospin conserving strong int.} ($\sim O(1)$)$|$ $\gg$ 
$|${\it radiative int.} ($\sim O(\sqrt{\alpha})$)$|$
$\gg$ $|${\it isospin non-conserving hadronic int.}
($\sim O(\alpha)$~\cite{Dalitz})$|$, 
%%%%%%%%%%%%%%%%%%%%%%%%%%%%%%%%%%%%%%%%%%%%%%%%%%%%%%%%%%%%%%%%%%%%%%%
where $\alpha$ is the fine structure constant.  
Such a state cannot be any ordinary $\{c\bar{s}\}$ meson. 
When it is assigned to an iso-triplet scalar 
$\hat{F}_I^+\sim\{[cn][\bar{s}\bar{n}]\}_{I=1}^+$, ($n=u,\,d$) meson, 
its narrow width is understood by a small overlap of color and spin wave functions 
(wfs.)~\cite{HT-isospin}, where the notation of tetra-quark states will be seen later. 
In addition, a recent lattice-QCD study on mass of the lowest-lying $I=0$ 
charm-strange ($C=S=1$) scalar-meson has reproduced~\cite{Namekawa} the 
measured one $m^{\rm exp}_{D_{s0}(2317)}$ of $D_{s0}^+(2317)$. 
This suggests that there exists an iso-singlet charm-strange scalar meson which is (approximately) degenerate with $D_{s0}^+(2317)$, and implies that it is a compact 
object but not any extended object like a loosely bound $DK$ molecule~\cite{BCL}, i.e.,  
there exist $D_{s0}^+(2317)$ as the iso-triplet $\hat{F}_I^+$ and its iso-singlet  
$\hat{F}^+\sim\{[cn][\bar{s}\bar{n}]\}_{I=0}^+$ partner whose indication has been 
observed in $D_s^{*+}\gamma$ channel~\cite{D_{s0}-Belle}. 
Regarding $X(3872)$, it is known that its $\pi^+\pi^-\psi$ decay proceeds through the 
intermediate $\rho^0\psi$ state~\cite{X-Belle,CDF-pipi}.  
Nevertheless, $X(3872)$ is considered to be an $I=0$ state, because its charged 
partners have not been observed~\cite{Babar-X-charged-partner}. 
In addition, its spin ($J$), parity ($P$) and $\mathcal{C}$-parity are given by 
$J^{P\mathcal{C}} = 1^{++}$~\cite{PDG13}.  
Thus, quantum numbers of $X(3872)$ are the same as those of the charmonia 
$\chi_{c1}(1P)$, $\chi_{c1}(2P)$, $\cdots$.   
If it were a charmonium, however, the ratio of decay rates 
%%%%%%%%%%%%%%%%%%%%%%%%%%%%%%%%%%%%%%%%%%%%%%%%%%%%%%%%%%%%%%%%%%%%%%%
$R({\gamma\psi/\pi^+\pi^-\psi})
=\Gamma(X(3872)\rightarrow \gamma\psi)/
\Gamma(X(3872)\rightarrow \pi^+\pi^-\psi)$  
%%%%%%%%%%%%%%%%%%%%%%%%%%%%%%%%%%%%%%%%%%%%%%%%%%%%%%%%%%%%%%%%%%%%%%%
would be much larger than unity~\cite{omega-rho-KT}, i.e., 
$R({\gamma\psi/\pi^+\pi^-\psi})_{c\bar{c}} \gg 1$, for the decay in the denominator 
would be suppressed because of the isospin non-conservation and the 
OZI-rule~\cite{OZI}. 
This result contradicts with the measurements, 
%%%%%%%%%%%%%%%%%%%%%%%%%%%%%%%%%%%%%%%%%%%%%%%%%%%%%%%%%%%%%%%%%%%%%%%%
\begin{equation} 
R({\gamma\psi/\pi^+\pi^-\psi})_{\rm exp}  
= 0.33 \pm 0.12\,\, ({\rm Babar})                     
\,\,{\rm and} 
\,\,0.22 \pm 0.09\,\, ({\rm Belle}),  
                                                                             \label{eq:gamma/pipi-exp}
\end{equation} 
%%%%%%%%%%%%%%%%%%%%%%%%%%%%%%%%%%%%%%%%%%%%%%%%%%%%%%%%%%%%%%%%%%%%%%%
which have been provided in \cite{Babar-X-gamma-psi} and obtained by compiling the  
data in \cite{Belle-X-gamma-psi}, respectively, and hence, it should be a multi-quark  
state. 
Here, an argument~\cite{prompt-X-theor} that the measured cross section for prompt  
$X(3872)$ production~\cite{prompt-X-CDF}  favors a compact object like a tetra-quark  
state over an extended one like a loosely bound meson-meson molecule should be 
noted. 
If it is the case, $X(3872)$ would be a tetra-quark meson. 
Concerning with $Z_c^{\pm,0}(3900)$, they cannot be charmonia and their neutral  
component $Z_c^0(3900)$ has an odd $\mathcal{C}$-parity, so that they can be 
interpreted as iso-triplet opposite $\mathcal{C}$ partners of $X(3872)$, as discussed  
before.  
As for $\hat{\delta}^{c0}(3200)$, it will be an $I = 1$ hidden-charm scalar meson. 
The simplest way to understand it is to assign it to a tetra-quark 
state~\cite{hidden-charm-scalar-KT} 
$\hat{\delta}^{c0}\sim\{[cn][\bar{c}\bar{n}]\}^0_{I=1}$. 
In this case, its mass is estimated to be $m_{\hat{\delta}^c}\simeq 3.3$ GeV, by using 
the quark counting in \cite{D_{s0}-KT} and taking $m^{\rm exp}_{D_{s0}^+(2317)}$ as the  
input data. 
The result is close to $m^{\rm exp}_{\hat{\delta}^c(3200)}\simeq 3.2$ GeV. 
In contrast, the diquark-antidiquark model~\cite{Maiani} and the unitarized chiral 
one~\cite{hidden-charm-scalar-mol} have predicted that the mass of the lowest   
hidden-charm scalar (called as $X_0$ in these papers) is $m_{X_0}\simeq 3.7$ GeV 
which is much higher  than $m^{\rm exp}_{\hat{\delta}^c(3200)}\simeq 3.2$ GeV. 
Therefore, $\hat{\delta}^{c0}(3200)$ will provide an important clue to select a realistic 
model of multi-quark states. 

We here review very briefly our tetra-quark model. 
Tetra-quark states can be classified into four groups, 
%%%%%%%%%%%%%%%%%%%%%%%%%%%%%%%%%%%%%%%%%%%%%%%%%%%%%%%%%%%%%%%%%%%%%%%
\begin{eqnarray} 
&&\hspace{-8mm} \{qq\bar q\bar q\} =  
[qq][\bar q\bar q] \oplus (qq)(\bar q\bar q)  
\oplus \{[qq](\bar q\bar q)\oplus (qq)[\bar q\bar q]\}, 
                                                   \label{eq:4-quark} 
\end{eqnarray} 
%%%%%%%%%%%%%%%%%%%%%%%%%%%%%%%%%%%%%%%%%%%%%%%%%%%%%%%%%%%%%%%%%%%%%%%
in accordance with difference of symmetry property of their flavor wfs., where 
parentheses and square brackets denote symmetry and anti-symmetry, 
respectively, of flavor wfs. under exchange of flavors between them~\cite{Jaffe}. 
Each term on the the right-hand-side (r.h.s.) of Eq.~(\ref{eq:4-quark}) is again 
classified into two groups~\cite{Jaffe} with 
%%%%%%%%%%%%%%%%%%%%%%%%%%%%%%%%%%%%%%%%%%%%%%%%%%%%%%%%%%%%%%%%%%%%%%%
${\bf{\bar{\bf{3}}}_c}\times{\bf{3}_c}$ and ${\bf{6}_c}\times {\bf{\bar{\bf{6}}_c}}$  
%%%%%%%%%%%%%%%%%%%%%%%%%%%%%%%%%%%%%%%%%%%%%%%%%%%%%%%%%%%%%%%%%%%%%%%
of the color $SU_c(3)$. 
Here, the former is taken as the lower-lying state in heavy mesons~\cite{D_{s0}-KT}.  
However, the second term $(qq)(\bar q\bar q)$ on r.h.s. of Eq.~(\ref{eq:4-quark}) is 
not considered in this note, because no signal of scalar $(K\pi)_{I=3/2}$ meson which 
can arise from $(qq)(\bar q\bar q)$ in the light flavor sector has been observed in a  
sufficiently wide enery region $\lesssim 1.8$ GeV~\cite{K-pi-3/2}.  
Regarding their $J^P$, the first term and the last two on the r.h.s. of 
Eq.~(\ref{eq:4-quark}) have $J^P=0^+$ and $1^+$, respectively, in the flavor symmetry 
limit, because $[qq]$ and $(qq)$ have $J^P=0^+$ and $1^+$, respectively, in the same 
limit. 
Nevertheless, the flavor symmetry is broken in the real world, so that $[qq]$ and $(qq)$ 
can have both of $J^P=0^+$ and $1^+$ in general, and hence each term on the r.h.s. 
of Eq.~(\ref{eq:4-quark}) can have all of $J^P=0^+,\,1^+$ and $2^+$. 
Along with this line, the diquark-antidiquark model~\cite{Maiani} in which a large 
breaking of isospin symmetry is assumed has assigned axial-vector mesons to 
$[qq][\bar{q}\bar{q}]$ states (which disappear in the flavor symmetry limit), and hence,  
$X(3872)$ to an element of $[cn][\bar{c}\bar{n}]$. 
As the result, it predicts $m_{X_0}\simeq 3.7$ GeV as the mass of the lowest 
hidden-charm scalar meson which is much higher than 
$m^{\rm exp}_{\hat{\delta}^c(3200)}$, as discussed before, and therefore, 
the diquark-antidiquark model fails to understand it. 
It is because $X(3872)$ has been assigned to an element of $[cn][\bar{c}\bar{n}]$ 
with $J^P=1^+$ (which disappears in the flavor symmetry limit). 

In contrast, we assign scalar and axial-vector tetra-quark mesons to different  
$[qq][\bar{q}\bar{q}]$ and $\{[qq](\bar{q}\bar{q})\oplus (qq)[\bar{q}\bar{q}]\}$, 
respectively, and therefore, $D_{s0}(2317)$ and $\hat{\delta}^c(3200)$ to 
$[cn][\bar{s}\bar{n}]_{I=1}$ and $[cn][\bar{c}\bar{n}]_{I=1}$, 
respectively~\cite{D_{s0}-KT,hidden-charm-scalar-KT}. 
In the $J^P=1^+$ mesons, $[qq](\bar q\bar q)$ and $(qq)[\bar q\bar q]$ are not 
eigenstates of $\mathcal{C}$-parity, so that they 
%a pair of $[cn](\bar{c}\bar{n})$ and $(cn)[\bar{c}\bar{n}]$, and a hidden-strangeness 
%pair of $[cs](\bar{c}\bar{s})$ and $(cs)[\bar{c}\bar{s}]$ 
mix with each other to form eigenstates of $\mathcal{C}$-parity. 
Therefore, we have pairs of hidden-charm axial-vector meson states with opposite 
$\mathcal{C}$-parities, i.e., 
%%%%%%%%%%%%%%%%%%%%%%%%%%%%%%%%%%%%%%%%%%%%%%%%%%%%%%%%%%%%%%%%%%%%%%%
$X(\pm) \sim \{[cn](\bar{c}\bar{n}) \pm (cn)[\bar{c}\bar{n}]\}_{I=0}$, 
$X_I(\pm) \sim \{[cn](\bar{c}\bar{n}) \pm (cn)[\bar{c}\bar{n}]\}_{I=1}$ and 
$X^s(\pm) \sim \{[cs](\bar{c}\bar{s}) \pm (cs)[\bar{c}\bar{s}]\}$, 
%%%%%%%%%%%%%%%%%%%%%%%%%%%%%%%%%%%%%%%%%%%%%%%%%%%%%%%%%%%%%%%%%%%%%%%
where $\pm$ denote the $\mathcal{C}$-properties ($\mathcal{C}$-parities of their  
neutral components), and the ideal mixing among tetra-quark states is always assumed 
in this note. 
(In this scheme, scalar and axial-vector mesons survive even in the flavor symmetry  
limit.) 
Thus, $X(3872)$ is assigned to $X(+)$ with $I=0$ and $\mathcal{C}=+$. 
In this case, the measured ratio Eq.~(\ref{eq:gamma/pipi-exp}) can be easily 
reproduced, by assuming that the isospin non-conserving 
$X(+) = X(3872)\rightarrow \pi^+\pi^-\psi$ decay in the denominator proceeds 
through the $\omega\rho^0$ mixing~\cite{omega-rho-KT} which plays important roles 
in the observed $\omega\rightarrow \pi\pi$ decay~\cite{PDG13} and the isospin 
non-conservation in nuclear forces~\cite{omega-rho-NF}. 
In addition, it has been argued~\cite{KT-partners-of-X} that $X_{I}(+)$ is considerably 
broad, when it is assumed (as an approximation) that its mass and spatial wf. are not 
very much different from those of $X(+) = X(3872)$, i.e., 
$m_{X_I(+)}\simeq m_{X(3872)}$ and the couplings of $X_{I}(+)$ to ordinary mesons 
(up to the Clebsch- Gordan coefficients arising from the color and spin degree of 
freedom) are not very far from those of $X(+)$. 
The above assumption seems to be natural, because these states belong to the same 
ideally-mixed $\{[cn](\bar{c}\bar{n}) + (cn)[\bar{c}\bar{n}]\}$ multiplet. 
Concerning with $X(-)$ and $X_I(-)$, the measured mass values of $Z_c(3900)$ 
%which is assigned to 
($=X_I(-)$ in the present scheme) are not very far from the measured 
one~\cite{PDG13} of  $X(3872)$ ($=X(+)$ in our model), i.e., 
%%%%%%%%%%%%%%%%%%%%%%%%%%%%%%%%%%%%%%%%%%%%%%%%%%%%%%%%%%%%%%%%%%%%%%%%
\begin{equation}
\frac{m_{X_I(-)} - m_{X(+)}}{m_{X(+)}} = 
\frac{m_{Z_c(3900)} - m_{X(3872)}}{m_{X(3872)}} 
\lesssim 1\,\,\%.                                                           \label{eq:mass-difference}
\end{equation}
%%%%%%%%%%%%%%%%%%%%%%%%%%%%%%%%%%%%%%%%%%%%%%%%%%%%%%%%%%%%%%%%%%%%%%%%
Therefore, we again assume that spatial wfs. of $X(-)$ and $X_I(-)$ are not very far 
from those of $X(+)$ and $X_I(+)$. 
Under this condition, it has been intuitively expected that $X(-)$ and $X_I(-)$ also 
are considerably broad~\cite{KT-partners-of-X}. 

To estimate numerically widths of $X(-)$ and $X_I(-)$, we first estimate  
phenomenologically the rate for $X(3872)\rightarrow D^0\bar{D}^{*0}$. 
We here identify~\cite{PDG13} $X(3872)$ with $X(3875)$ which was observed in the 
%%%%%%%%%%%%%%%%%%%%%%%%%%%%%%%%%%%%%%%%%%%%%%%%%%%%%%%%%%%%%%%%%%%%%%%%
$(D^0\bar{D}^{*0} + c.c.\rightarrow)\,\, D^0\bar{D}^{0}\pi^0$ 
%%%%%%%%%%%%%%%%%%%%%%%%%%%%%%%%%%%%%%%%%%%%%%%%%%%%%%%%%%%%%%%%%%%%%%%%
channel. 
Assuming that the total rate $\Gamma_{X(3872)}$ is approximately saturated as  
%%%%%%%%%%%%%%%%%%%%%%%%%%%%%%%%%%%%%%%%%%%%%%%%%%%%%%%%%%%%%%%%%%%%%%%%
\begin{equation}
\Gamma_{X(3872)} \simeq \Gamma(X(3872) \rightarrow \pi^+\pi^-\psi)
+ \Gamma(X(3872) \rightarrow \pi^+\pi^-\pi^0\psi) 
+ \Gamma(X(3875)\rightarrow D^0\bar{D}^{*0} + c.c.), 
\end{equation}
%%%%%%%%%%%%%%%%%%%%%%%%%%%%%%%%%%%%%%%%%%%%%%%%%%%%%%%%%%%%%%%%%%%%%%%%
and taking the measured ratios of rates~\cite{Choi}, 
%%%%%%%%%%%%%%%%%%%%%%%%%%%%%%%%%%%%%%%%%%%%%%%%%%%%%%%%%%%%%%%%%%%%%%%%
$[{\Gamma(X(3875)\rightarrow D^0\bar{D}^{*0} + c.c.)}/
{\Gamma(X(3872)\rightarrow \pi^+\pi^-\psi)}]_{\rm exp}= 9.5\pm 3.1$ 
%%%%%%%%%%%%%%%%%%%%%%%%%%%%%%%%%%%%%%%%%%%%%%%%%%%%%%%%%%%%%%%%%%%%%%%%
and $[{\Gamma(X(3872)\rightarrow \pi^+\pi^-\pi^0\psi)}/
{\Gamma(X(3872)\rightarrow \pi^+\pi^-\psi)}]_{\rm exp}= 0.8\pm 0.3$, 
%%%%%%%%%%%%%%%%%%%%%%%%%%%%%%%%%%%%%%%%%%%%%%%%%%%%%%%%%%%%%%%%%%%%%%%%
in addition to the measured width~\cite{Belle-D^0D^{*0}bar} 
%%%%%%%%%%%%%%%%%%%%%%%%%%%%%%%%%%%%%%%%%%%%%%%%%%%%%%%%%%%%%%%%%%%%%%%%
$\Gamma_{X(3875)} = (3.9^{+2.8+0.2}_{-1.4-1.1})$ MeV,  
%%%%%%%%%%%%%%%%%%%%%%%%%%%%%%%%%%%%%%%%%%%%%%%%%%%%%%%%%%%%%%%%%%%%%%%%
we obtain 
%%%%%%%%%%%%%%%%%%%%%%%%%%%%%%%%%%%%%%%%%%%%%%%%%%%%%%%%%%%%%%%%%%%%%%%%
\begin{equation}
\Gamma(X(3872)\rightarrow D^0\bar{D}^{*0}) 
\sim (0.3 - 1.5)\,\,{\rm MeV}.                                 \label{eq:width-X}
\end{equation}
%%%%%%%%%%%%%%%%%%%%%%%%%%%%%%%%%%%%%%%%%%%%%%%%%%%%%%%%%%%%%%%%%%%%%%%%
This result is consistent with an independent estimate~\cite{Renga}, 
${\Gamma(X(3872)\rightarrow D^0\bar{D}^{*0})_{\rm Renga} \sim 1}$ MeV. 

Next, we write the rate for the $X(+)=X(3872)\rightarrow D^0\bar{D}^{*0}$ decay 
(as an example of $1^+\rightarrow 0^- + 1^-$ decays) as 
%%%%%%%%%%%%%%%%%%%%%%%%%%%%%%%%%%%%%%%%%%%%%%%%%%%%%%%%%%%%%%%%%%%%%%%%
\begin{equation}
\displaystyle{
\Gamma(X(+)\rightarrow D^0\bar{D}^{*0}) 
= \frac{|g_{X(+)D^0\bar{D}^{*0}}|^2}{24\pi m_{X(+)}^2}{p_D}
\biggl\{
2+ \frac{(m_{X(+)}^2 - m_{D^0}^2 + m_{D^{*0}}^2)^2}
{4 m_{X(+)}^2m_{D^{*0}}^2}\biggr\}},                             \label{eq:rate-D^0antiD^{*0}}   
\end{equation}
%%%%%%%%%%%%%%%%%%%%%%%%%%%%%%%%%%%%%%%%%%%%%%%%%%%%%%%%%%%%%%%%%%%%%%%%
where $g_{X(+)D^0\bar{D}^{*0}}$ denotes the $X(+)D^0\bar{D}^{*0}$ coupling strength 
and $p_D$ is the size of the center-of-mass momentum of $D^0$ in the final state. 
To study two- and three-body decays (through quasi-two-body intermediate states) 
of $X(-)$ and $X_I(\pm)$, we remember that we can decompose each of them into a 
sum of products of two $\{q\bar{q}\}$ pairs (and hence two ordinary mesons) as 
Eqs.~(10) $-$ (13) in \cite{KT-partners-of-X}. 
Then, under the above assumptions (as an approximation) on spatial wfs. of  
$X(+)=X(3872)$ and its hidden-charm partners $X_I(\pm)$ in addition to $X(-)$,  
ratios of their coupling-strengths to ordinary mesons to $g_{X(+)D^0\bar{D}^{*0}}$ 
are given by ratios of corresponding (Clebsch-Gordan) coefficients arising from color 
and spin degrees of freedom in the decompositions. 
In this way, rates for OZI-rule-allowed decays of $X_I(\pm)$ and $X(-)$ are very  
crudely estimated as listed in Table I, where $m_{X(-)}\simeq m_{X_I(-)}=m_{Z_c(3900)}$ 
has been assumed and Eq.~(\ref{eq:width-X}) has been taken as the input data. 
Besides, the $\eta\eta'$ mixing with the mixing angle~\cite{PDG10} 
$\theta_P\simeq -20^\circ$ in $X(-)\rightarrow\eta\psi$ and the broad width of  
$\rho^0$ meson in decays through intermediate $\rho^0$ states have been taken into  
account. 
%%%%%%%%%%%%%%%%%%%%%%%%%%%%%%%%%%%%%%%%%%%%%%%%%%%%%%%%%%%%%%%%%%%%%%%%
\begin{center}%\hspace{-30mm}
\begin{table}[t]       %
\begin{quote}
%\caption{
Table~I. 
Rates for OZI-rule-allowed two- and three-body decays of hidden-charm partners of 
$X(3872)$ are listed, where it is assumed that the spatial wave functions of ${X(\pm)}$  
and ${X_I(\pm)}$ are nearly equal to each other. 
The rate $\Gamma(X(+)\rightarrow D^0\bar{D}^{*0})\sim (0.3 - 1.5)$ MeV which is 
given in the text is taken as the input data.   
%}
\end{quote} \vspace{2mm}
\begin{tabular}{l|l|l|l}
\hline
\hspace{10mm}Decay & \hspace{5mm}Rate (MeV) 
& \hspace{10mm}Decay & \hspace{5mm}Rate (MeV)\\
\hline
{${X_I^0(+)\rightarrow \rho^0\psi\rightarrow\pi^+\pi^-\psi}$}
& \hspace{1.5mm} $\sim 20\hspace{2.5mm} - 200$  \hspace{2mm} (${\ddagger}$)  
&
{${X_I^0(+)\rightarrow D^0\bar{D}^{*0}}$}
& \hspace{1.5mm} $\sim\hspace{1.8mm}0.3 - \hspace{3mm}1.5$ \\
\hline
{${X\hspace{1.5mm}(-)\rightarrow \eta_c\omega}$} 
& \hspace{1.5mm} $\sim \hspace{1.5mm}9 \hspace{2.5mm}- \hspace{1.5mm}45$ 
& {${X\hspace{1.5mm}(-)\rightarrow \eta\psi}$}
& \hspace{1.5mm}  $\sim\hspace{1.5mm} 7 \hspace{2.5mm}- \hspace{2mm}35$ 
\hspace{2mm} (${\ast}$) \\
\hline
{${X_I^0(-)\rightarrow \pi^0\psi}$}
& \hspace{1.5mm}  $\sim \hspace{0mm}15 \hspace{2.5mm}- \hspace{1.2mm}75$ 
& {${X_I^0(-)\rightarrow \eta_c\rho^0\rightarrow \eta_c\pi^+\pi^-}$}
& \hspace{1.5mm} $\sim\hspace{1.5mm}6\hspace{2.5mm} - \hspace{2.0mm}30$
\\   
\hline
\end{tabular} \vspace{2mm}\\
\hspace{-0mm}($\ddagger$) Ref.~\cite{KT-partners-of-X}.   
\hspace{5mm}($\ast$) The ${\eta\eta'}$ mixing 
with the mixing angle ${\theta_P \simeq -20^\circ}$~\cite{PDG10}. 
\end{table}\vspace{-4mm}
\end{center}
%%%%%%%%%%%%%%%%%%%%%%%%%%%%%%%%%%%%%%%%%%%%%%%%%%%%%%%%%%%%%%%%%%%%%%%

Full width $\Gamma_{X^0_I(-)}$ is approximately given by a sum of rates for dominant 
two- and three-body decays, i.e., 
$\Gamma_{X^0_I(-)}\simeq \Gamma(X^0_I(-)\rightarrow \pi^0\psi) 
+ \Gamma(X_I^0(-)\rightarrow \eta_c\rho^0\rightarrow \eta_c\pi^+\pi^-)
\sim (20 - 100)$ MeV, and in a similar way, $\Gamma_{X(-)}\sim (15 - 80)$ MeV. 
The results imply that the widths of $X_I(-)$ and $X(-)$ are considerably broad as 
intuitively expected. 
Therefore, their detection in $B$ decays will require much higher statistics than those 
to observe $X(3872)$. 
Fortunately, however, $Z^{\pm,0}_c(3900)$ ($= X_I(-)$ in the present scheme) have 
been observed in 
%%%%%%%%%%%%%%%%%%%%%%%%%%%%%%%%%%%%%%%%%%%%%%%%%%%%%%%%%%%%%%%%%%%%%%%%
$e^+e^-\rightarrow Y(4260)\rightarrow \pi^+\pi^-\psi$, and 
$e^+e^-\rightarrow \psi(4160)\rightarrow \pi^+\pi^-\psi$ and $\pi^0\pi^0\psi$,  
%%%%%%%%%%%%%%%%%%%%%%%%%%%%%%%%%%%%%%%%%%%%%%%%%%%%%%%%%%%%%%%%%%%%%%%%
as discussed before. 
It should be noted that our estimate of $\Gamma_{X_I(-)}$ is consistent with the 
measured widths of $Z_c^{\pm,0}(3900)$, and therefore, our assumption on spatial wfs. 
of $X(\pm)$ and $X_I(\pm)$ mesons seems to be feasible.  

As seen above, our tetra-quark interpretation of $D_{s0}^+(2317)$, 
$\hat{\delta}^{c0}(3200)$ and $X(3872)$ seems to be favored by experiments. 
In addition, the measured mass and width of $Z_c^{\pm,0}(3900)$ are consistent with 
our predictions. 
However, to establish our tetra-quark interpretation, observation of states which 
have been predicted in our model would be needed. 
In this sense, confirmation of existence of the $I=0$ partner 
$\hat{F}^+\sim\{[cn][\bar{s}\bar{n}]\}^+_{I=0}$ of 
$D_{s0}^{+}(2317)=\hat{F}_I^+\sim\{[cn][\bar{s}\bar{n}]\}_{I=1}^{+}$ 
would take priority, because its indication has already been observed in the  
$D_s^{*+}\gamma$ channel from $B$ decays~\cite{D_{s0}-Belle}. 
(It should be noted that production of such a state in $e^+e^-$ annihilation is  
suppressed. 
This can be understood by considering their production in a framework of minimal 
$\{q\bar{q }\}$ pair creation~\cite{production-D_{s0}-KT}.)  
In addition, our tetra-quark model has predicted~\cite{D_{s0}-KT,Exotic-QN} existence 
of scalar and axial-vector mesons with exotic quantum numbers. 
Therefore, their observation is one of important options to establish our tetra-quark 
interpretation. 
Neutral and doubly charged partners $D_{s0}^0(2317)$ and $D_{s0}^{++}(2317)$ of 
$D_{s0}^+(2317)$ are candidates of such states, where 
$D_{s0}^{0,+,++}(2317) = \hat{F}_I^{0,+,++}\sim\{[cn][\bar{s}\bar{n}]\}_{I=1}^{0,+,++}$ 
in the  present scheme~\cite{D_{s0}-KT}. 
Although they have not been observed in inclusive $e^+e^-$ 
annihilation~\cite{Babar-D_{s0}-charged-partners}, it does not necessarily imply their 
non-existence. 
In fact, it is expected that their production is {\it suppressed} in the inclusive 
$e^+e^-$ annihilation~\cite{production-D_{s0}-KT}. 
Therefore, they should be searched in $B$ decays, because branching fractions for 
their productions in $B$ decays have been estimated to be large enough to observe them~\cite{production-D_{s0}-KT}, i.e.,   
%%%%%%%%%%%%%%%%%%%%%%%%%%%%%%%%%%%%%%%%%%%%%%%%%%%%%%%%%%%%%%%%%%%%%%%%
$Br(B_u^+\rightarrow D^{(*)-}D_{s0}^{++}(2317))
\sim Br(B_d^0\rightarrow \bar{D}^{(*)0}D_{s0}^0(2317))  
\sim (10^{-4} - 10^{-3})$. 
%%%%%%%%%%%%%%%%%%%%%%%%%%%%%%%%%%%%%%%%%%%%%%%%%%%%%%%%%%%%%%%%%%%%%%%%

Double-charm ($C=2$) axial-vector mesons are  
%%%%%%%%%%%%%%%%%%%%%%%%%%%%%%%%%%%%%%%%%%%%%%%%%%%%%%%%%%%%%%%%%%%%%%%%
$H_{A{cc}}^+\sim (cc)[\bar u\bar d]$ and 
$K_{A{cc}}^{+,++}\sim\{(cc)[\bar n\bar s]\}^{+,++}$ 
%%%%%%%%%%%%%%%%%%%%%%%%%%%%%%%%%%%%%%%%%%%%%%%%%%%%%%%%%%%%%%%%%%%%%%%%
in our model~\cite{Exotic-QN}.                                                       
Their detection will be another option to establish our interpretation. 
Their masses have been very crudely estimated as 
%%%%%%%%%%%%%%%%%%%%%%%%%%%%%%%%%%%%%%%%%%%%%%%%%%%%%%%%%%%%%%%%%%%%%%%%
$m_{H_{A{cc}}}\simeq 3.87$ GeV and $m_{K_{A{cc}}}\simeq 3.97$ GeV 
%%%%%%%%%%%%%%%%%%%%%%%%%%%%%%%%%%%%%%%%%%%%%%%%%%%%%%%%%%%%%%%%%%%%%%%%
by using the same quark counting as before, where 
%%%%%%%%%%%%%%%%%%%%%%%%%%%%%%%%%%%%%%%%%%%%%%%%%%%%%%%%%%%%%%%%%%%%%%%%
$m_{(cc)[\bar{u}\bar{d}]}\simeq m_{(cn)[\bar{c}\bar{n}]}
\simeq m_{[cn](\bar{c}\bar{n})}\simeq m_{X(3872)}$ 
%%%%%%%%%%%%%%%%%%%%%%%%%%%%%%%%%%%%%%%%%%%%%%%%%%%%%%%%%%%%%%%%%%%%%%%%
has been assumed. 
These results are close to thresholds of possible OZI-rule-allowed two- and 
three-body decays of these mesons.  
Because deviations between the estimated masses and thresholds under consideration  
might be smaller than uncertainties involved in their estimated mass values, we here 
get rid of taking literally these values. 
%%%%%%%%%%%%%%%%%%%%%%%%%%%%%%%%%%%%%%%%%%%%%%%%%%%%%%%%%%%%%%%%%%%%%%%%
Here it should be noted that the above $C=2$ meson states might correspond to 
a part of $T_{cc}$'s in \cite{Yasui-Lee}, although their mass values estimated in these 
two different models are not necessarily agree with each other. 
If $T_{cc}$'s are stable against their OZI-rule-allowed strong decays as discussed 
in \cite{Yasui-Lee}, they should be very narrow and could be observed as sharp peaks 
in $DD_{(s)}\gamma$ channels in $B_c$ decays and in inclusive $e^+e^-$ annihilation 
if their production rate is sufficiently high~\cite{Hyodo}. 
%%%%%%%%%%%%%%%%%%%%%%%%%%%%%%%%%%%%%%%%%%%%%%%%%%%%%%%%%%%%%%%%%%%%%%%%
On the other hand, in our case, it is not very clear whether their OZI-rule-allowed 
strong decays are kinematically allowed or not, because the estimated mass values 
of these mesons are very close to corresponding thresholds of these decays. 
% OZI-rule-allowed decays. 
Even though their true masses are a little bit higher than the thresholds of these 
decays, however, they would be narrow because of their small phase space volume. 
(We here do not estimate their widths, because it is difficult to get a definite result 
under the present condition.) 
Regarding $K_{A{cc}}^{+,++}$, therefore, we expect intuitively that they will be observed 
as narrow peaks in $DD_s^+\gamma$ (and $DD_s^+\pi$ if kinematically allowed) 
channel(s) in $B_c^+$ decays.  
It is because branching fractions for $B_c^+$ decays producing them, which have been  
estimated very crudely as~\cite{Exotic-QN} 
%%%%%%%%%%%%%%%%%%%%%%%%%%%%%%%%%%%%%%%%%%%%%%%%%%%%%%%%%%%%%%%%%%%%%%%%
$Br(B_c^+\rightarrow \{\bar{D}^{(*)}K_{A{cc}}\}^+) \sim (10^{-4} - 10^{-3})$, 
%%%%%%%%%%%%%%%%%%%%%%%%%%%%%%%%%%%%%%%%%%%%%%%%%%%%%%%%%%%%%%%%%%%%%%%%
might be large enough to observe them. 
In contrast, observation of $H_{A{cc}}^+$ in $B_c^+$ decays would be not very easy, 
because its production in $B_c^+$ decays is CKM suppressed~\cite{Exotic-QN}. 
Here, it should be noted that observation of double-charm mesons will exclude the 
diquark-antidiquark model. 

Observation of exotic $C=-S=1$ meson states is an additional option. 
However, the scalar $\hat{E}^0\sim [cs][\bar{u}\bar{d}]$ decays only through weak 
interactions~\cite{D_{s0}-KT}, so that its detection might not be easy. 
Regarding axial-vector tetra-quark mesons $E^0_{A{(cs)}}\sim(cs)[\bar{u}\bar{d}]$ 
and $E_{A{[cs]}}^{\pm,0}\sim\{[cs](\bar{n}\bar{n})\}^{\pm,0}$, their masses have been 
very crudely estimated~\cite{Exotic-QN} to be $2.97$ GeV by using the same quark  
counting as the above. 
The above value is sufficiently higher than thresholds of their possible 
OZI-rule-allowed two-body-decays. 
Therefore, it can be intuitively expected that they are considerably broad, because of  
large phase space volume. 
To study numerically their decay rates, we here decompose $E^0_{A{(cs)}}$ as in  
\cite{KT-partners-of-X}, 
%%%%%%%%%%%%%%%%%%%%%%%%%%%%%%%%%%%%%%%%%%%%%%%%%%%%%%%%%%%%%%%%%%%%%%%%
\begin{eqnarray}
&&    
E^0_{A{(cs)}}    
=\frac{1}{4\sqrt{3}}\biggl\{
D^{*+}K^- + D^+K^{*-} - D^{*0}\bar{K}^0 - D^0\bar{K}^{*0} 
+ 
\bar{K}^{*0}D^0 + \bar{K}^0D^{*0} - K^{*-}D^+ - K^-D^{*+} 
\biggr\} + \cdots,                                                       \label{eq:decomp-E{(cs)}}
\end{eqnarray}
%%%%%%%%%%%%%%%%%%%%%%%%%%%%%%%%%%%%%%%%%%%%%%%%%%%%%%%%%%%%%%%%%%%%%%%%
where $\cdots$ denotes a sum of products of color octet $\{q\bar{q}\}_{\bm{8_c}}$ pairs. 
$E^{\pm,0}_{A{[cs]}}$ also can be decomposed in the same way. 
Here, we list only the decomposition of $E^+_{A{[cs]}}$ to save space,  
%%%%%%%%%%%%%%%%%%%%%%%%%%%%%%%%%%%%%%%%%%%%%%%%%%%%%%%%%%%%%%%%%%%%%%%%
\begin{equation}
\hspace{-13mm}
E^+_{A{[cs]}}   
=\frac{1}{2\sqrt{6}}\biggl\{
D^{*+}\bar{K}^0 + D^+\bar{K}^{*0} - \bar{K}^{*0}D^+ - \bar{K}^{0}D^{*+} 
\biggr\} + \cdots                                           \label{eq:decomp-E{[cs]}}
\end{equation}
%%%%%%%%%%%%%%%%%%%%%%%%%%%%%%%%%%%%%%%%%%%%%%%%%%%%%%%%%%%%%%%%%%%%%%%%

Rates for the  $E^0_{A{(cs)}}\,({\rm or}\, E_{A{[cs]}})\rightarrow \bar{K}D^{*}$ 
[and $D\bar{K}^{*}$] decays which are considered as their dominant ones are obtained 
by replacing $X(+)$, $D^0$ and $\bar{D}^{*0}$ in Eq.~(\ref{eq:rate-D^0antiD^{*0}}) in 
terms of $E^0_{A{(cs)}}$ (or  $E_{A{[cs]}}$), $\bar{K}$[$D$] and $D^{*}$[$\bar{K}^*$], 
respectively. 
In this way,  the ratios of rates 
%%%%%%%%%%%%%%%%%%%%%%%%%%%%%%%%%%%%%%%%%%%%%%%%%%%%%%%%%%%%%%%%%%%%%%%%
$\Gamma(E^0_{A{(cs)}}\rightarrow \bar{K}D^*[D\bar{K}^*])
/\Gamma(X(+)\rightarrow D^0\bar{D}^{*0})$
%%%%%%%%%%%%%%%%%%%%%%%%%%%%%%%%%%%%%%%%%%%%%%%%%%%%%%%%%%%%%%%%%%%%%%%%
are given by the ratios of coupling strengths  
$|g_{E^0_{A{(cs)}}\bar{K}D^*[D\bar{K}^*]}/g_{X(+)D^0\bar{D}^{*0}}|^2$.   
The latter ratios are provided by the ratios of the Clebsch-Gordan coefficients in the 
above Eq.~(\ref{eq:decomp-E{(cs)}}) to the ones of Eq.~(10) in \cite{KT-partners-of-X},  
under the assumption (as an approximation) that the spatial wf. overlaps among  
$E^0_{A{(cs)}}$, $\bar{K}\,[D]$ and $D^{*0}\,[\bar{K}^*]$ are not very much different from  
that of $X(+)$, $D^0$ and $\bar{D}^{*0}$. 
In the same way, rates for two-body decays of $E^+_{A{[cs]}}$ also can be estimated 
as listed in Table II, where the rate in Eq.(\ref{eq:width-X}) has been taken as the input  
data. 
However, the results are still very preliminary, because they include large uncertainties. 
%%%%%%%%%%%%%%%%%%%%%%%%%%%%%%%%%%%%%%%%%%%%%%%%%%%%%%%%%%%%%%%%%%%%%%%
\begin{center}\hspace{-30mm}
\begin{table}[t]
\begin{quote}
%\caption{
Table~II.  
Rates for two-body decays of $E^0_{A{(cs)}}$ and $E^+_{A{[cs]}}$  are listed, where  
the rate $\Gamma(X(+)\rightarrow D^0\bar{D}^{*0})\sim (0.3 - 1.5)$ MeV 
which is given in the text is taken as the input data.  
%}
\end{quote}
\begin{tabular}{l|l|l|l}
\hline
\hspace{5mm}Decay & \hspace{0mm}Rate (MeV)
&  \hspace{5mm}Decay & \hspace{0mm}Rate (MeV) 
\\   
\hline
{${E^0_{A{(cs)}}\rightarrow \bar{K}^0D^{*0}}$}
& \hspace{1.5mm} $\sim \hspace{1.5mm} 6 - 30$  %\hspace{8mm} (${\ddagger}$)  
%\\\hline
&
{${E^0_{A{(cs)}}\rightarrow K^-{D}^{*+}}$}
& \hspace{1.5mm} $\sim \hspace{1.5mm}  6 - 30$ 
\\
\hline
{${E^0_{A{(cs)}}\rightarrow \bar{K}^{*0}D^0}$} 
& \hspace{1.5mm} $\sim \hspace{1.5mm}  5 - 25$ 
%\\  \hline
&
{${E^0_{A{(cs)}}\rightarrow {K}^{*-}D^+}$}
& \hspace{1.5mm}  $\sim \hspace{1.5mm} 5 - 25$ 
%\hspace{8mm} (${\ast}$) 
\\
\hline
{${E_{A{[cs]}}^+\rightarrow \bar{K}^0D^{*+}}$}
& \hspace{1.5mm}  $\sim 12 - 65$ 
%\\  \hline
&
{${E_{A{[cs]}}^+\hspace{0.5mm}\rightarrow \bar{K}^{*0}D^+}$}
& \hspace{1.5mm} $\sim 10 - 50$
\\   
\hline
\end{tabular} 
\end{table} \vspace{-8mm}\\
\end{center}
%%%%%%%%%%%%%%%%%%%%%%%%%%%%%%%%%%%%%%%%%%%%%%%%%%%%%%%%%%%%%%%%%%%%%%%

Full widths $\Gamma_{E^0_{A{(cs)}}}$ and $\Gamma_{E^+_{A{[cs]}}}$ 
are approximately given by a sum of the rates for two-body decays listed in Table~II, 
i.e.,  $\Gamma_{E^0_{A{(cs)}}}\simeq \Gamma_{E^+_{A{[cs]}}}\sim (20 - 100)$ MeV. 
From this, it is seen that they are much broader than $X(3872)$. 
On the other hand, branching fractions for $B$ decays which produce $E^0_{A_{(cs)}}$ 
and $E^0_{A{[cs]}}$ have been estimated~\cite{Exotic-QN} to be 
%%%%%%%%%%%%%%%%%%%%%%%%%%%%%%%%%%%%%%%%%%%%%%%%%%%%%%%%%%%%%%%%%%%%%%%
$Br(\bar{B}\rightarrow \bar{D}E^0_{A{(cs)}})\sim 
Br(\bar{B}\rightarrow \bar{D}E^0_{A{[cs]}})\sim (10^{-4} - 10^{-3})$. 
%%%%%%%%%%%%%%%%%%%%%%%%%%%%%%%%%%%%%%%%%%%%%%%%%%%%%%%%%%%%%%%%%%%%%%%
Production of $E^{\pm}_{A{[cs]}}$ is also described by the same type of quark-line 
diagrams as those decribing $E^{0}_{A{[cs]}}$ production, so that the branching fractions  
for the $B$ decays producing them also are expected to be of the same order of  
magnitude, i.e., $Br(\bar{B}\rightarrow \bar{D}E^{\pm}_{A{[cs]}})\sim (10^{-4} - 10^{-3})$. 
These results are not very much different from those of $B$ decays producing  
$D_{s0}^+(2317)$ and $X(3872)$. 
Therefore, detection of $E^0_{A{(cs)}}$ and $E^{\pm,0}_{A{[cs]}}$ in $B$ decays will 
require much higher statistics than those to observe $D_{s0}^+(2317)$ and $X(3872)$. 

In summary, we have presented our tetra-quark interpretation of $D_{s0}^+(2317)$, 
$X(3872)$ and $\hat{\delta}^{c0}(3200)$ (the $\eta\pi^0$ peak around 3.2 GeV 
indicated in $\gamma\gamma$ collision), and have pointed out that confirmation of 
$\hat{\delta}^c(3200)$ can be a clue to select a realistic model of multi-quark mesons. 
Next, we have studied decay properties of tetra-quark partners of $X(3872)$. 
However, the hidden-charm and exotic $C=-S=1$ partners are expected to be broad,  
and therefore, detection of them in $B$ decays will require much higher statistics 
than the ones to observe $X(3872)$. 
Therefore, to search for them at the present experimental accuracy, some other 
processes should be studied. 
Fortunately, $Z^{\pm,0}_c(3900)$ ($= X^{\pm,0}_I(-)$ in the present scheme) have been  
observed in exclusive $e^+e^-\rightarrow \pi^+\pi^-\psi$ annihilations. 
%Although numerical study of 
However, decay property of $X^s(\pm)$ is left as one of our future subjects. 
%, $X^s(+)$ is intuitively expected to be narrow because of a small phase space volume. 
Regarding $K^{+,++}_{A{cc}}$ with $C=2$ and $S=1$,  they have been expected to be  
narrow. 
If their production rates are sufficiently high, therefore, they will be observed as 
narrow peaks in  ($DD_s^+\pi$ and) $DD_s^+\gamma$ channels in inclusive $e^+e^-$  
annihilation. 
If not, however, they will be observed as narrow peaks in the same channels of the 
$B_c^+$ decays 
$B_c^+\rightarrow \{\bar{D}^{(*)}K_{A{cc}}\}^+\rightarrow (\{\bar{D}^{(*)}(DD_s^+\pi)\}^+$ 
{and}) $\{\bar{D}^{(*)}(DD_s^+\gamma)\}^+$. 

In addition, confirmation of the iso-singlet partner of $D_{s0}^{+}(2317)$ in the 
$D_s^{*+}\gamma$ channel, and observation of $D_{s0}^{0,++}(2317)$ 
in $D_s^+\pi^\pm$ channels from $B$ decays are awaited. 

%%%%%%%%%%%%%%%%%%%%%%%%%%%%%%%%%%%%%%%%%%%%%%%%%%%%%%%%%%%%%%%%%%%%%%%
\section*{Acknowledgments}    
The author would like to thank Professor H.~Kunitomo for careful reading of the 
manuscript. 
%%%%%%%%%%%%%%%%%%%%%%%%%%%%%%%%%%%%%%%%%%%%%%%%%%%%%%%%%%%%%%%%%%%%%%%

%%%%%%%%%%%%%%%%%%%%%%%%%%%%%%%%%

%\end{references}
%%%%%%%%%%%%%%%%%%%%%%%%%%%%%%%%%%%%%%%%%%%%%%
\end{document}